\documentstyle[11pt,paspconf,psfig]{article}
 \begin{document}

\title{High Redshift HI 21cm Absorption toward Red Quasars}
\author{C.L. Carilli}
\affil{National Radio Astronomy Observatory}
\author{Karl M. Menten}
\affil{Max-Planck-Institut f{\"u}r Radioastronomie}
\author{C.P. Moore}
\affil{Kapteyn Institute}

\begin{abstract}

We have searched for redshifted absorption in the 21 cm line of neutral
hydrogen toward `red quasars', which are extragalactic radio sources
with a steep spectral drop between optical and infrared wavelengths.
The success rate for detecting HI 21cm absorption 
toward a representative sample of such sources is 80$\%$. This compares
to the much lower success rate of 11$\%$ for detecting HI 21cm
absorption associated with optically selected Mg II absorption line
systems. The large neutral hydrogen column densities seen 
toward  red quasars 
supports the
hypothesis that these sources are reddened by dust, as opposed to 
having an intrinsically red spectrum due to the AGN emission mechanism. 
The lower limits to the  spin temperatures for the neutral hydrogen are
between 50 K and 1000~K, assuming a  Galactic dust-to-gas ratio. 
 
We consider the question of biases in
optically selected samples of quasars due to dust obscuration. 
The data on the red quasar sub-sample
support the models  of Fall \&\ Pei
for dust obscuration by damped Ly$\alpha$ absorption line systems,
and suggest that:
(i) there may be a significant, but not dominant, population of 
quasars missing from optically selected samples due to
dust obscuration, perhaps as high as 20$\%$ at the POSS limit
for an optical sample 
with a redshift distribution similar to the 1 Jy, flat spectrum quasar
sample, and (ii) optically
selected samples may miss about half the high column density
quasar absorption line systems.

\end{abstract}


\index{Active Galactic Nucleus}  \index{HI 21cm Absorption}  
\index{Quasar Absorption Lines}  
\index{Red Quasars}  \index{HI Spin Temperature}  
\index{Dust Absorption}  
\index{Interstellar Medium}  
\index{WSRT}  \index{NRAO 140ft Telescope}  \index{0108+388}  
\index{0218+357}  \index{0414+055}  
\index{0500+019}  \index{Circumnuclear Gas}  \index{1413+135} 
\index{1504+377}  
\index{Gravitational Lensing}  \index{1830-211}  \index{Extinction}

\section{Introduction}

Quasar absorption line systems provide a distance-independent means of
studying baryonic matter cosmologically distributed throughout the
universe. The neutral hydrogen column density distribution
function for these systems  can be characterized roughly by a single
power-law between  10$^{13}$ cm$^{-2}$ and 10$^{22}$
cm$^{-2}$, with an index of
$-$1.5. (Tytler 1987).
The implication is that absorption line systems with neutral
hydrogen column  densities greater than 10$^{21}$ cm$^{-2}$ are 
very rare:  for every 1000 absorption line systems with
N(HI) $<$ 10$^{14}$ cm$^{-2}$ there is one system with N(HI) $>$
10$^{21}$ cm$^{-2}$. 
Although the  highest column density systems are
rare, study of these systems is critical to our understanding of the 
high redshift universe for a number of reasons, including:

\begin{itemize}

\item  Most of the neutral baryonic material at high redshift 
may reside in the highest column density systems (Lanzetta et al. 1995).

\item The very high column  systems are thought to be absorption by 
gas associated with  the pre-cursors of present day disk galaxies
(Wolfe et al. 1986).

\item Obscuration by dust associated with the high column density
systems may affect the observed statistics of quasar populations at high
redshifts, and perhaps more significantly, the observed statistics of the 
cosmic evolution of the neutral baryon density (Fall \&\ Pei 1995). 

\item Perhaps most importantly in the cases of  radio loud
background sources, the high column density systems 
allow studies of the HI 21cm line in
absorption, and in the more extreme cases, of associated molecular 
absorption at high redshift. 

\end{itemize}

Spectroscopic imaging at radio wavelengths
of  molecular and HI 21cm absorption line systems 
opens a new window into the physics of these systems, including 
detailed studies of astrochemistry at high redshift, and new methods
for determining basic 
cosmological parameters such as the evolution of the temperature of
the microwave background radiation and the deuterium abundance.
These systems probe the dense, pre-star-forming
interstellar medium (ISM) in galaxies at substantial look-back
times. Background radio continuum sources typically show spatial
structure on scales ranging from parsecs to kilo-parsecs. Radio
interferometers  allow for direct spectroscopic imaging of the spatial
structure in the absorbing  gas over this entire range of spatial scales. 
Also, radio spectroscopy allows for study of the velocity structure in
the absorption lines down to 10 m s$^{-1}$. 

\section{Red Quasars}

Over the last few years we have developed an effective  technique for
discovering very high column density quasar absorption line systems by
searching for redshifted HI 21cm absorption toward
flat spectrum radio sources with faint, red optical
counterparts. The physical basis for selecting these `red quasars' for
radio absorption  searches is the hypothesis that these sources are
red due to dust obscuration. Dust obscuration leads to a natural bias
against finding the highest column density absorbers using optical
techniques in two ways: (i) sources may drop out of optical magnitude
limited samples due to dust obscuration, and (ii) follow-up 
optical spectroscopy typically requires fairly bright, blue objects
(Heisler \&\ Ostriker 1988,
Fall \&\ Pei 1993, Webster et al. 1995, Shaver et al. 1996).
Hence, radio spectroscopy may provide the {\sl only method} for
studying the very highest column density quasar absorption line
systems, as has long been the case for studies of the early stages of
star formation in Galactic dark molecular clouds. 

In this paper we present a summary of recent detections of
redshifted neutral hydrogen 21cm absorption toward red quasars
using the new UHF system at the Westerbork Synthesis Radio
Telescope (WSRT), and the NRAO 140ft telescope in Green Bank WV
(see Carilli et al. 1997, 1998). These observations are intended to 
address a number of issues. First, we want to address the 
question whether some AGNs have intrinsically red,
rather than dust-reddened, spectra. 
Second is the question of the fraction of quasars missing from optical
samples due to dust obscuration, as well as the fraction of missing
high column density quasar absorption line systems.
Third, we wish to enlarge the
sample of dusty, high column density absorption line systems 
in order to study the dense ISM in 
galaxies at significant look-back times.
An earlier review of redshifted HI 21cm absorption line systems can be
found in Carilli (1995). More recent results using the new UHF
system at the WSRT can be found in Vermeulen et al. and Lane et al. 
(this volume). 

The resulting sample of very high column density absorption line
systems detected in the radio  via HI 21cm absorption and/or molecular
absorption, falls into two classes: (i) associated absorbers,
ie. absorption by gas 
in the host galaxy of the quasar, and (ii) cosmologically intervening
absorbers. For the associated absorbers the gas can be either
`circumnuclear material' directly associated with the active galactic
nucleus (AGN), or gas in the general ISM of the AGN host galaxy. For the
cosmologically intervening systems the interesting trend has arisen
that  many of these systems are gravitational lenses. In retrospect
the association 
of lensing with very high column density absorbers may not be surprising,
since high column density absorption requires a
fairly small impact parameter on the intervening galaxy. The surprise
has been that many gravitational lenses are gas rich systems, as
opposed to the original naive hypothesis that most lenses would be
gas poor elliptical galaxies. 

Parameters for the radio continuum sources, and for Gaussian models
fit to  the HI absorption lines, are given in Table 1.  The reference
redshift corresponding to zero velocity is given in each figure, and in
the table, as z$_0$. 

\section{Individual Sources}

\subsection{0108+388}
 
The radio source 0108+388 is located at the center of a narrow emission 
line galaxy at z = 0.670  (Stickel et al. 1996a,
Stanghellini et al. 1993). The radio source shows 
twin-jets on 3 mas scales, plus extended structure on large scales (Taylor,
Readhead, \&\ Pearson 1996).
This source also shows an inverted spectrum at low frequency, perhaps
indicating free-free absorption. Optical images of the field show a
very red, `slightly asymmetric' galaxy, perhaps a face-on
spiral (Stanghellini et al. 1993).  
The lack of a point source in R band images, and the presence of 
a compact I band source, have led Stanghellini et al. (1993)
to propose `the existence of nuclear obscuration' toward 0108+388. 

A strong HI 21cm absorption line is detected at z = 0.66847
with a width of 100 km  s$^{-1}$ and a peak optical depth of
0.44 (Carilli et al. 1998). 
The integrated HI column density for this system
is 81$\times$10$^{18}$$\times({{T_s}\over{f}})$ cm$^{-2}$,
where T$_s$ is the spin
temperature of the gas and $f$ is the HI covering factor (Figure 1). 
Using the VLBI structure of the radio source, and the high opacity of
the HI line,  Carilli et al. (1998) estimate a 
lower limit to the absorbing cloud size of
about 6 mas, or 36 pc (H$_o$ = 65 km s$^{-1}$ Mpc$^{-1}$ and q$_o$
= 0.5). 

The  velocity width for the HI 21cm absorption line toward
0108+388 is comparable to  linewidths
seen for systems associated with nearby Seyfert
galaxies and other low redshift AGN (Pedlar et al., this volume;
Conway, this volume). High resolution imaging of the
absorption toward a number of nearby AGN indicates that 
the high velocity dispersion gas is local to the AGN, perhaps
in a circumnuclear torus or ring with size of order 10 pc,  and
that the velocity width reflects the torus dynamics (Pedlar et al.,
and Conway this volume).  

\begin{figure}
\psfig{figure=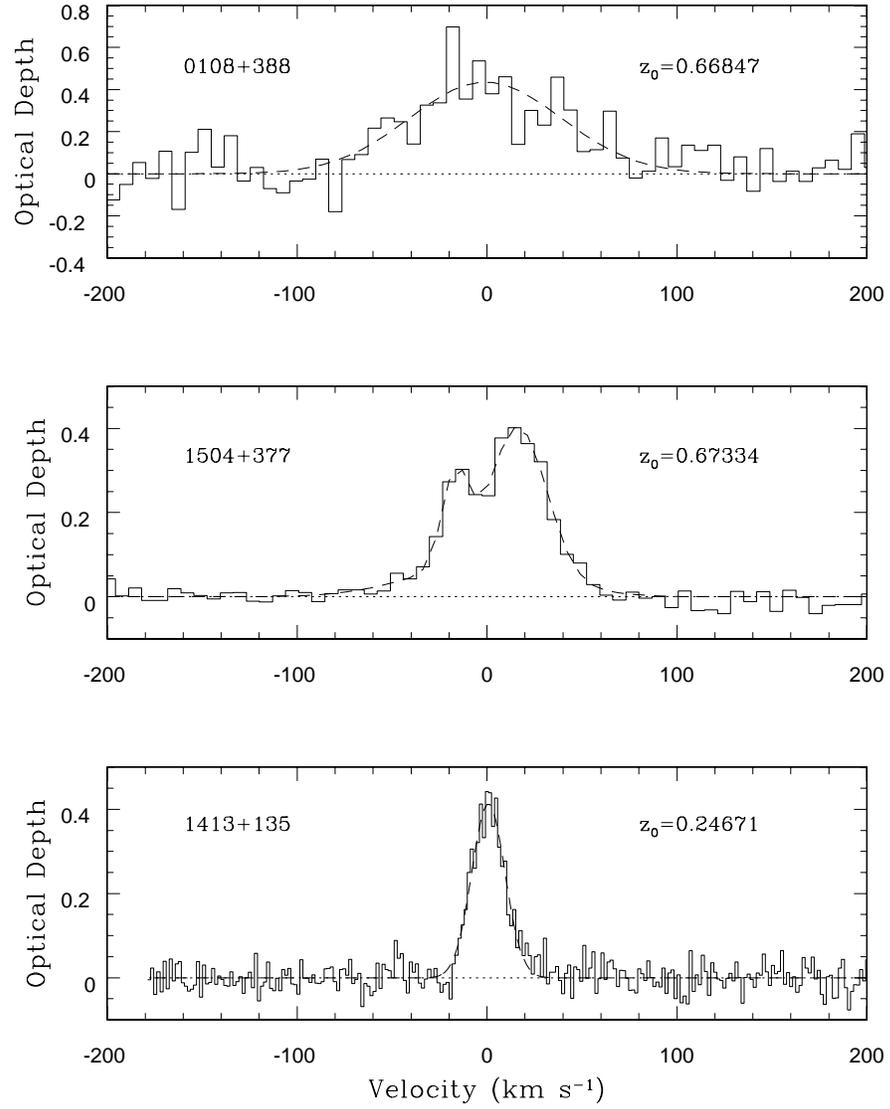,width=12cm}
\caption{Spectra of redshifted HI 21cm absorption toward three red
quasars. These three systems are likely to be absorption by 
gas associated with the host galaxy of the AGN.
The sources and redshifts are listed in each frame. The
redshifts  listed correspond to zero velocity.
The parameters for the radio source and the 
Gaussian model fit, are given in Table 1.}
\end{figure}

\subsection{0218+357}

The radio source 0218+357 is the `smallest Einstein ring', with a
radio ring with diameter of $0{\rlap.}{''}37$ plus two  point sources, one at
the center of the ring and the second on the ring. Differential
reddening in the optical toward the two compact sources has led to the
hypothesis of a dusty gravitational lens (O'Dea et al. 1992, Patnaik
et al. 1993). HI 21cm absorption has been detected toward 0218+357 
by the lensing galaxy at z = 0.6847 (Carilli, Rupen, \&\ Yanny 1993),
with N(HI) = 3$\times$10$^{18}$$\times({{T_s}\over{f}})$ cm$^{-2}$ 
(Figure 2). Subsequent imaging spectroscopy has
revealed molecular absorption at z = 0.6847 toward  the redder of the two
compact radio components (Wiklind \&\ Combes 1995 and this volume; Menten \&\
Reid 1996). Lens models predict a velocity dispersion for the lensing
galaxy  of 100 km s$^{-1}$, suggesting lensing by the bulge of a
spiral  galaxy. The HI 21cm line has a FWHM = 43 km s$^{-1}$,
suggesting that the disk of the lensing galaxy may be fairly close to
the sky plane (Carilli et al. 1993). 

\subsection{0414+055}

The radio source 0414+055 is an `Einstein quad' with a
maximum component separation of 1$''$. 
Optical and near IR imaging by
Lawrence et al. (1995) indicates that the
optical components are 
heavily reddened by dust. The source redshift is 2.636.
The redshift of the lens remains unknown. Moore, Carilli, \&\
Menten (1999) have detected HI 21cm absorption at z = 2.636 toward
0414+055 with N(HI) = 7.5$\times$10$^{18}$$\times({{T_s}\over{f}})$
cm$^{-2}$ (Figure 3). The large velocity dispersion
of this gas (FWHM $\approx$ 300 km s$^{-1}$) suggests absorption by
circumnuclear material. The observed HI 21cm absorption 
implies that much of the reddening  is by gas in the AGN host
galaxy. However, differential colors for the optical components may
require additional reddening by dust in the lens (McLeod et al. 1998).

\subsection{0500+019}

The radio source 0500+019 is unresolved at cm wavelengths at resolutions
ranging from a few mas to 
a few arcseconds (Taylor \&\ Perley 1996). 
A 2.2 $\mu$m image of this source reveals a point source
with faint emission extending 3$''$ to the south, while an R band image shows
an inclined  disk galaxy with a major axis of about 4$''$ (Stickel et
al. 1996b).   There is no evidence for a point source in the R band image,
although the galaxy brightness profile is asymmetric, peaking about 
1$''$ to the north of center along the major axis of the galaxy, 
coincident within the errors with the position of the K band point source.
The galaxy redshift is  z = 0.5834$\pm$0.0014, but there is 
a narrow emission line at 0.6354 $\mu$m 
which cannot be reasonably identified with the galaxy (Stickel et al. 1996b).
This unidentified emission line, the off-center
location of the 2.2 $\mu$m point source relative to the galaxy center,
and the asymmetric galaxy profile in the R band image, lead Stickel et
al. to suggest that the quasar may be a background source at z $>$ 0.5834,
and that the red color of the quasar is due to dust in the foreground
galaxy. 

HI 21cm absorption has been detected toward 0500+019 at z = 0.58472,
with N(HI) = 7$\times$10$^{18}$$\times({{T_s}\over{f}})$ cm$^{-2}$
(Carilli et al. 1998). The observed spectrum is fit reasonably
well by two Gaussians (Figure
3). The total HI linewidth is about 140 km s$^{-1}$. This
broad profile toward 0500+019 could be used to
argue for absorption by circumnuclear material.
However, there are the additional puzzles that 
the HI redshift is offset from the optical redshift by  240
km s$^{-1}$, plus the offset between the K band point source and the
galaxy center and the unidentified optical emission line. 
A possible model to explain all these data is that the quasar is a
background source which projects off-center from the z = 0.5834 galaxy. 
The large velocity offset of the HI line relative to the 
galaxy, and the large velocity dispersion, would then be 
explained by galactic rotation. For instance, assuming  a  galaxy 
rotation velocity of 200 km sec$^{-1}$, and the fact that
the line-of-sight to the quasar cut through the edge-on disk halfway out
along the major axis, the expected linewidth due to differential
rotation would be about 100 km s$^{-1}$.

\subsection{1413+135}

The flat spectrum radio source 1413+135 is located at the center of
an edge on spiral galaxy at z = 0.247 (Perlman et al. 1996), and the
radio AGN is situated behind the galaxy's dust lane (McHardy et
al. 1994).  Carilli, Perlman, \&\ Stocke (1992) have detected HI 21cm
absorption at z = 0.247 with 
N(HI) = 17$\times$10$^{18}$$\times({{T_s}\over{f}})$
cm$^{-2}$ (Figure 1). The HI 21cm absorption line is fairly narrow
(FWHM $\approx$ 20 km s$^{-1}$), and very narrow molecular absorption
(FWHM $\approx$ 1 km s$^{-1}$) has been detected at the redshift of
the HI line (Wiklind \&\ Combes 1994 and this volume). The narrow velocity
width of the absorption toward 1413+135 suggests
that it does not occur in circumnuclear gas, but in the general
ISM of the AGN host galaxy. Perlman et al. propose that 1413+135 is a
young radio loud AGN (age $\le$ 10$^4$ years) behind a giant molecular
cloud in the disk of the host galaxy. 

\begin{figure}
\psfig{figure=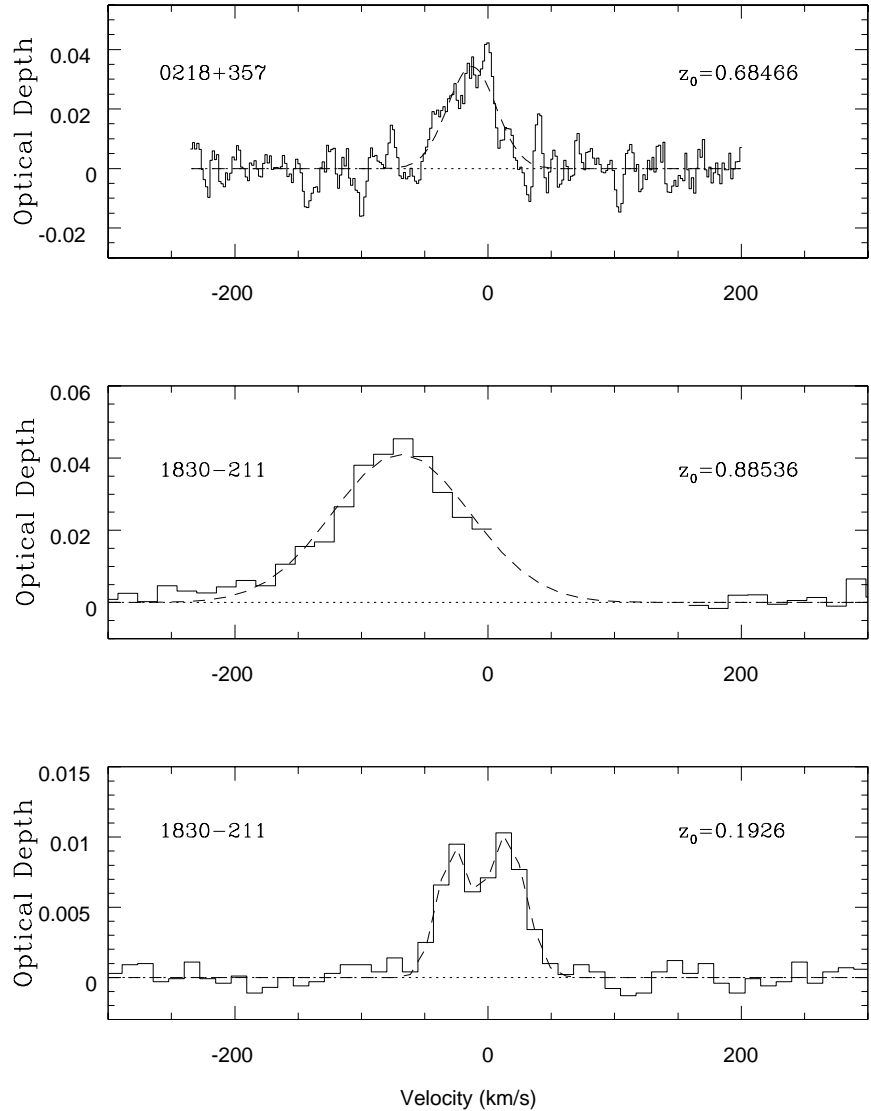,width=12cm}
\caption{Spectra of redshifted HI 21cm absorption toward the Einstein
ring radio sources 1830-211 and 0218+357. The
redshifts  listed correspond to zero velocity. 
These systems are due to cosmologically intervening gas.
The parameters for the Gaussian model fit, and the source continuum
flux densities, are given in Table 1.
Note that the spectrum of 1830-211  is corrupted 
by terrestrial interference velocity = +100 km s$^{-1}$ in this
spectrum. A second peak in HI
21cm absorption has been detected at this velocity  (see Chengalur, de
Bruyn, \&\ Narasimha, this volume).
}
\end{figure}

\subsection{1504+377}

The flat spectrum radio source 1504+377
is located at the center of an inclined disk
galaxy with a moderate-excitation narrow emission line spectrum at z = 
0.674$\pm$0.001. 
The source shows evidence for flaring of its  near IR intensity,
although optical images of 1504+377  show no indication of a bright AGN 
(Stickel et al. 1996a).
Radio spectroscopic observations of 1504+377 have revealed
strong HI 21cm absorption (Figure 1), and absorption by a number of
molecular species, at the redshift of the parent galaxy (Wiklind \&\
Combes 1996a, Carilli et al. 1998). The HI profile is
reasonably fit by three Gaussians, with  a total N(HI) = 
45$\times$10$^{18}$$\times({{T_s}\over{f}})$ cm$^{-2}$.
The broad absorption line in
this case again argues for absorption by circumnuclear material 
(Wiklind \&\ Combes 1996a, Carilli et al. 1997, 1998). 

The molecular absorption toward 1504+377 also shows a narrow feature
at z = 0.67150 (linewidth $\le$ 10 km s$^{-1}$), or 330 km s$^{-1}$ relative to
the broad line. No HI 21cm absorption is detected at 
the redshift of this narrow molecular feature. This and 
the velocity difference between the broad and narrow absorption
components in 1504+377 is discussed by Carilli et al. (1997, 1998). 

\subsection{1830-211}

The radio source 1830$-$211 is the `brightest Einstein ring', with a
ring of 1$''$ diameter and two point sources located on opposite sides
of the ring. The source remains unidentified in the optical, but there
is a possible detection in the near IR (Frye et al. this
volume). Strong molecular and HI 21cm absorption has been detected toward
1830$-$211 at z = 0.886 (Wiklind \&\ Combes 1996b and this volume; Frye
et al. 1997 and this volume; Menten, Carilli, \&\ Reid, this volume;
Chengalur et al., this volume). The HI 21cm
absorption is broad, with two components separated by 146 km
s$^{-1}$. Spectroscopic imaging of the molecular absorption 
shows strong absorption toward the SW radio component at z = 0.88582,
and weaker absorption toward the NE component at $-$146 km s$^{-1}$
relative to the strong lines. This velocity separation is consistent
with the dynamics for the lensing galaxy predicted by lens models
(Wiklind \&\ Combes 1998 and this volume; Menten et al. this volume). 

A second HI 21cm absorption system has been discovered toward
1830$-$211 at z = 0.1926 with N(HI) = 
1.2$\times$10$^{18}$$\times({{T_s}\over{f}})$ cm$^{-2}$.
No molecular absorption has been detected at this redshift (Wiklind
\&\ Combes this volume). This second absorption component complicates
the lens models for 1830$-$211, implying that this source is a
compound gravitational lens (Lovell et al. 1997). 

\begin{figure}
\psfig{figure=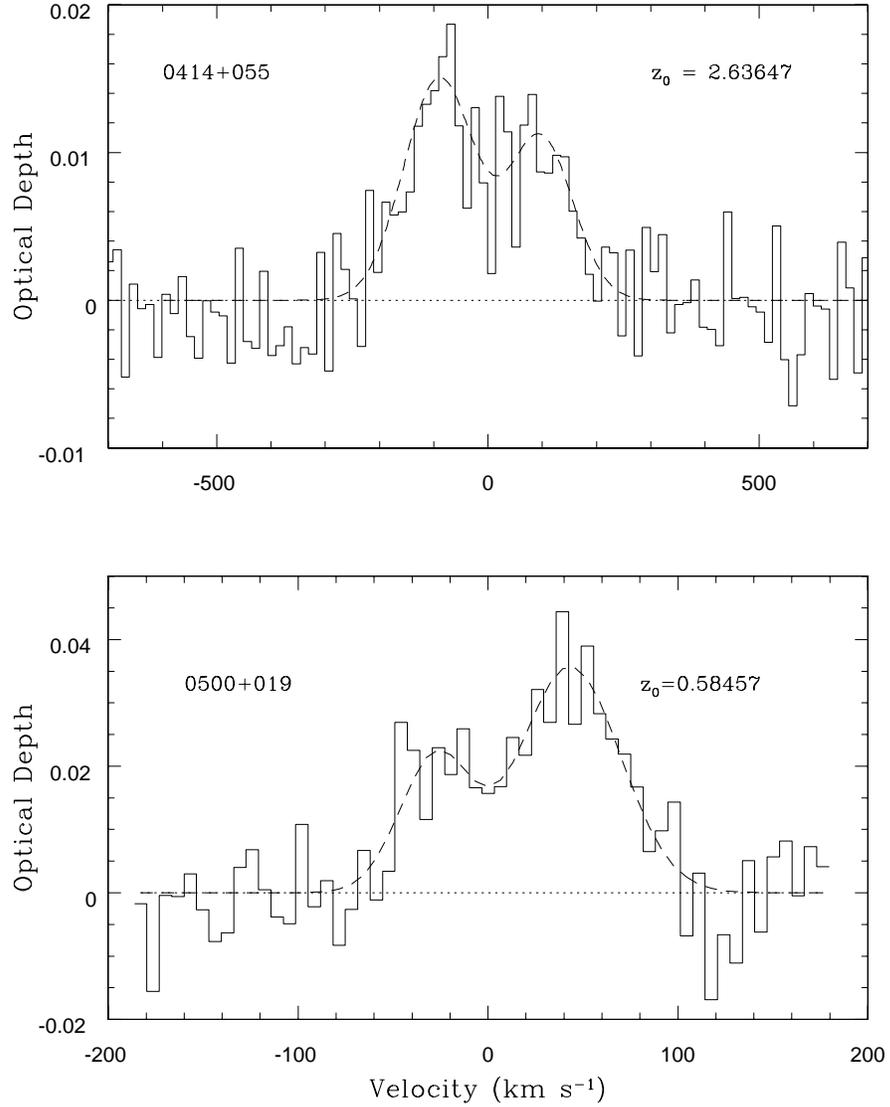,width=12cm}
\caption{Spectra of redshifted HI 21cm absorption toward the `Einstein
Quad'  radio source 0414+055, and toward 0500+019. The
redshifts  listed correspond to zero velocity. 
The absorption toward 0414+055 is by gas associated with the AGN.
For 0500+019 it is unclear whether the gas is associated with the AGN, or
cosmologically intervening. Note the different velocity scales for
the two frames. 
The parameters for the Gaussian model fit, and the source continuum
flux densities, are given in Table 1.}
\end{figure}

\begin{table}
\begin{center}
Table 1. ~~Gaussian Model Parameters

\scriptsize
\begin{tabular}{cccrcrrc}
\tableline
\tableline
\noalign{\vskip 0.07in}
~ & S$_\nu$ & z$_{0}$ & Velocity & Int. Opt. Depth & Opt. Depth & FWHM
& N(HI) \\ 
\tableline
\noalign{\vskip 0.07in}
~ & Jy & ~ & km s$^{-1}$ &  km s$^{-1}$ & ~ & km s$^{-1}$ & 
$\times$${T_s}\over{f}$10$^{18}$cm$^{-2}$ \\ 
\tableline
\noalign{\vskip 0.07in}
0108+388 & 0.18 & 0.66847 & 0$\pm$4 & 46$\pm$7 & 0.44$\pm$0.04
& 94$\pm$10 & 81$\pm$12 \\
\tableline
\noalign{\vskip 0.07in}
0218+357 & 1.8 & 0.68466 & $-$13$\pm$7 & 1.7$\pm$0.3 &
0.034$\pm$0.005 &  45$\pm$4 & 3.0$\pm$0.5 \\
\tableline
\noalign{\vskip 0.07in}
0414+055 & 3.3 & 2.63647 & 93$\pm$12  & 1.6$\pm$0.4 & 0.011$\pm$0.001 &
140$\pm$33 & 3.0$\pm$0.8 \\
0414+055 & ~ & ~ & $-$94$\pm$12 & 2.4$\pm$0.5 & 0.015$\pm$0.001 & 
154$\pm$28 & 4.5$\pm$0.9 \\
\tableline
\noalign{\vskip 0.07in}
0500+019 & 1.6 & 0.58457 & 43$\pm$3 & 2.5$\pm$0.3 & 0.036$\pm$0.003
& 62$\pm$7 & 4.5$\pm$0.5 \\
0500+019 & ~ & ~ & $-$27$\pm$5 & 1.4$\pm$0.3 
& 0.027$\pm$0.003 & 45$\pm$9  & 2.5$\pm$0.4 \\
\tableline
\noalign{\vskip 0.07in}
1413+135 & 1.1 & 0.24671 & 0$\pm$2 & 9.6$\pm$0.6 & 0.41$\pm$0.02 &
21$\pm$1 &  17$\pm$1 \\
\tableline
\noalign{\vskip 0.07in}
1504+377 & 1.0 & 0.67150 & 325$\pm$19 & 7.0$\pm$4.3 & 
0.073$\pm$0.045 & 85$\pm$22 & 12.8$\pm$8  \\
1504+377 & ~ & ~ & 313$\pm$1 & 4.0$\pm$0.6 & 0.22$\pm$0.03
& 16$\pm$2  & 7.3$\pm$1 \\
1504+377 & ~ & ~ & 347$\pm$1 & 13.4$\pm$4 & 0.34$\pm$0.08
& 34$\pm$5 & 24$\pm$7 \\ 
\tableline
\noalign{\vskip 0.07in}
1830$-$211 & 11 & 0.88536 & $-$141$\pm$3 & 5.8$\pm$0.4 & 0.041$\pm$0.002 &
127$\pm$6 & 10.1$\pm$0.7 \\
1830$-$211 & 12 & 0.19260 & $-$28$\pm$2 & 0.28$\pm$0.05 & 0.009$\pm$0.001 &
30$\pm$5 &  0.53$\pm$0.1 \\
1830$-$211 & 12 & 0.19260 & 15$\pm$2 & 0.36$\pm$0.05 & 0.010$\pm$0.001 &
35$\pm$5 &  0.68$\pm$0.1 \\
\tableline
\end{tabular}
\end{center}
\end{table}

\section{Discussion}
\subsection{The Statistics of Red Quasars}

We  have searched for HI 21cm absorption 
toward  a representative sub-sample of five red quasars from the
complete sample of Stickel et al. (1996a).
(Note that 1830$-$211,
1413$-$135, and 0414+055 are not in the Stickel et al.  red quasar
sub-sample due to various selection criteria.)
We have detected high column density
absorption systems in four of five sources (Carilli et al. 1998).
This 80$\%$ detection rate is significantly
higher than in optically selected samples:
searches for HI 21cm absorption associated with optically
selected MgII absorption line systems result in two detections out
of 18 systems  (Lane et al. this volume).

The average value of the radio-to-optical spectral index, 
$\alpha_{rad}^{opt}$, for optically identified, flat
spectrum radio loud quasars is:~  $\alpha_{rad}^{opt}$ = $-0.6$
(Condon et al. 1983), while the values for the Stickel et al. (1996a)
red quasar sub-sample are: $\alpha_{rad}^{opt}$ $<$ $-0.9$. 
On the other hand, the radio-to-near IR spectral indices 
for most of the red quasar sub-sample are: $\alpha_{rad}^{IR}$
$\ge$  $-0.8$, implying significant steepening of the spectra
between the near IR and the optical:~
$\alpha_{IR}^{opt} \ll -1$ (Stickel et al. 1996a). 
A rough lower limit to the required extinction can be derived by
comparing the observed optical flux density with that predicted by
extrapolating the radio-to-IR spectrum into the optical. This is a
lower limit due to confusion by emission from
stars in the galaxy with which the absorbing gas is 
associated. Values of the rest frame visual
extinction, A$_V$, range from: ~A$_V$ $\ge$ 2 for 
0108+388 and 0500+019, to A$_V$ $\ge$ 3 for 0218+357, and  A$_V$ $\ge$
7 for 1504+377.  
Using the observed value of N(HI) derived from the HI 21cm absorption lines
leads to  lower limits to the spin temperatures of about 50 K for
0108+388, 300 K for 1504+377, 500 K for 0500+019, and 1000 K for 0218+357,
assuming a Galactic dust-to-gas ratio, and $f~=~1$. 

Carilli et al. (1998) consider in detail the question of biases in
optically selected samples of quasars due to dust obscuration, 
using the  statistics of red quasars and 
the high detection rate of HI 21cm absorption toward 
red quasars. They find that the fraction of quasars
`missing' from  optically selected samples
due to dust obscuration is between 6$\%$ and 20$\%$, and that
the fraction of high column density absorption systems missing from 
optically selected samples of quasars is between 40$\%$ and 70$\%$, where
`optically selected sample' means a sample of quasars selected from a 
moderately deep optical survey such as the POSS, with a redshift
distribution comparable to the 1 Jy quasar sample.   

Fall and Pei (1993) have 
presented detailed models of the numbers of quasars missing from 
optically selected samples due to obscuration by dust associated with
high column density quasar absorption line systems as a function of
redshift. Using the redshift distribution for the 1 Jy sample, the
models of Fall and Pei  
predict that between 2$\%$ and 12$\%$ of the sources would be missing
from such a sample due to dust obscuration. 
The comparative numbers for the fraction of missing high column
density quasar absorption line systems 
from the models of  Fall and Pei (1993) are between 30$\%$ and
60$\%$. In both cases (missing quasar
fraction and missing high column absorber fraction), the models
of Fall and Pei agree reasonably well with the data on the 1 Jy
sample.

We should emphasize that there are a number of significant 
uncertainties in the statistics presented above. 
First is simply the small number of red quasars searched thus far for
redshifted HI 21cm absorption. 
Second are the possible biases introduced
by the fact that, due to
practical observational limitations, our absorption searches were
limited to the specific redshifts of the parent galaxies of the 
absorbing clouds, as determined from emission lines seen
in  deep optical spectra (Stickel et al. 1996a). 
It is possible that we have under-estimated the number
of high column density systems by not searching over the full
redshift range in each case.
Conversely, this selection criterion might skew the results toward
absorbers at the redshift of the quasar host galaxy,
which perhaps could bias the statistics in the
opposite sense. 
And third is the fact that the fairly high radio flux density 
limit of the 1 Jy sample
biases the sample toward lower redshift quasars. The redshift
distribution of the 1 Jy quasar
sample peaks at z $\approx$ 1, while quasar samples derived from
radio surveys with 
lower flux density limits peak at z $\approx$ 2, close to the 
redshift peak found for optically selected quasar samples (Shaver et
al. 1996). 
While the redshift distribution is used explicitly
in the calculations above, unbiased 
searches for redshifted HI 21cm absorption
toward a sample of red quasars selected from fainter radio catalogs
would make for a fairer comparison with the optical data. 

Overall, the statistics presented above should not be considered
rigorous, but only representative. Still, the red quasar data
support the basic conclusions of Fall and Pei (1993):
(i) there may be a significant, but not dominant, population of 
quasars missing from optically selected samples due to
obscuration, perhaps as high as 20$\%$ at the POSS limit
for an optical sample 
with a redshift distribution similar to the 1 Jy, flat spectrum quasar
sample, and (ii) optically
selected samples may miss about half the high column density
quasar absorption line systems.
A final point made by Fall and Pei is that the strongest bias in  optical
samples  may be against the high column density systems with
high metalicities and  dust-to-gas ratios. 
Such systems might be the most interesting 
from the perspective of follow-up radio studies 
of the dense, pre-star forming ISM in galaxies at significant look-back
times. The data presented herein suggests that 
searching for redshifted HI 21cm absorption toward radio-loud
red quasars may be
an effective method for circumventing this bias.

\acknowledgments{The National Radio Astronomy Observatory is a 
facility of the National Science Foundation
operated under cooperative agreement by Associated 
Universities, Inc.}

\end{document}